\documentclass[reprint,prc,groupedaddress,showpacs,preprintnumbers,amsmath,amssymb]{revtex4-1}
\usepackage{graphicx}
\usepackage{dcolumn}
\usepackage{amsmath}
\usepackage{times}
\usepackage{color}
\usepackage{float}
\usepackage[section]{placeins}

\begin{document}

\title{ Refinement of the $n-\alpha$ and $p-\alpha$ fish-bone potential }

\author{E.\ Smith}
\author{R.\ Woodhouse}
\author{Z.\ Papp}
\affiliation{ Department of Physics and Astronomy,
California State University Long Beach, Long Beach, California, USA }

\date{\today}

\begin{abstract}
The fishbone potential of composite particles 
simulates the Pauli effect by nonlocal terms. We determine the $n-\alpha$ and $p-\alpha$ fish-bone potential
by simultaneously fitting to the experimental phase shifts.
We found that with a double Gaussian parametrization of the local potential can describe the $n-\alpha$ and $p-\alpha$ phase shifts for all partial waves. 
\end{abstract}


\maketitle

\section{Introduction}

In nature, we can hardly find true elementary particles. Basically most of them are composite particles made of 
even more elementary constituents. These constituents are fermions.  Fermions obey the Pauli principle, i.e.\ they cannot 
occupy the same quantum state. On the other hand, when we try to place two 
composite particles at the same location we try 
to force that the constituents to occupy the same quantum state. The quantum system tries to prevent this and we 
observe the phenomena of Pauli blocking.

The simplest way to model the Pauli blocking is to use a local repulsive short range potential. This suppresses 
the wave function at short distances and reduce the probability of finding the constituents there.  
Since the Pauli blocking depends strongly on quantum numbers
these type of potentials exhibit a strong dependence on partial waves.

In fact, the Pauli blocking is a restriction on the Hilbert space. For composite particles the available Hilbert space is not 
the same as for structureless particles. Those states, which are occupied by the constituents, are
absent or suppressed in the relative motion.

There are several models for composite particle interaction, based on cluster model,  which, more or less,
follow this philosophy.
Probably the most elaborated is the fish-bone model by Schmid \cite{schmid1,schmid2}. In the fish-bone potential
the fully Pauli forbidden states are removed from the Hilbert space. This model also uses the concept of partly Pauli 
forbidden states, whose contribution is suppressed. As a result, the fish-bone potential is a combination of local and 
non-local terms. The structure of non-locality is determined by the internal structure of the constituents and the local 
potential is to be determined from a fitting procedure. 

In the fish-bone model we put all the information about the internal structure of the composite particle in their mutual 
interaction and we hope that we will achieve some simplification in parameters. Unfortunately, this was only partly true. In the case of the 
$\alpha-\alpha$ fish-bone potential, the parametrization of Ref.\ \cite{kircher} provided a good description of the 
two-body experimental phase shift, but it 
needed quite a sizable three-body potential to get the correct binding energy for the
three-$\alpha$ system \cite{oryu}.

In a recent study we reexamined the $\alpha-\alpha$ fish-bone potential  \cite{our3a}. 
We fitted the fish-bone potential to 
the $l=0$, $l=2$ and $l=4$ experimental phase shifts, the $l=0$ two-$\alpha$ resonant state and the low energy 
three-$\alpha$ ground 
and excited states. We found that a single Gaussian term in the local part of the fish-bone potential provides 
a reasonably good description of all these data. There is 
no need for any explicit angular momentum dependence and there is no need for a three-body potential. 
If the composite structure of the particles is 
properly built into the non-locality of the interaction, the fitted local part of the potential became really simple.

The $n-\alpha$ and $p-\alpha$ fish-bone potential has been determined in Ref.\ \cite{hahn}.
For the local part of the potential, the authors adopted a Wood-Saxon shape with a spin-orbit term.
Although they observed an excellent fit to the experimental phase shifts, the agreement was achieved by using 
independent parameters for each partial wave and different strength parameters for the spin-orbit term for
partial waves $\mbox{}^{2}P_{1/2}$ and $\mbox{}^{2}P_{3/2}$.

However, some of the parameters are rather close to each other. So, it may be possible to find a better parametrization 
of this potential which is, in the spirit of the fish-bone model, simple and does not have explicit angular momentum 
dependence. Maybe, the  angular momentum dependence we observe in nature, entirely comes from the 
composite structure of the alpha particles.

In Sec.\ II we outline the fish-bone optical model. In Sec.\ III we specialize it to the $n-\alpha$ and $p-\alpha$ case
and present our results. In Sec.\ IV we summarize and draw some conclusions.

\section{The fish-bone model for composite particle interaction}

The fish-bone model  has been derived from the resonating group model. 
In the resonating group model the total wave function is an antisymmetrized 
product of the cluster $|\Phi\rangle$ and the inter-cluster $|\chi\rangle$ relative states
\begin{equation}
|\Psi \rangle = |\{{\cal A} \Phi \chi \}\rangle.
\end{equation}
The state $|\Phi\rangle$  describes the internal properties of the clusters, including the spin and isospin structure. 
The unknown relative motion state $|\chi\rangle$ is determined from the variational ansatz
\begin{equation}
\langle \Phi \delta \chi | {\cal A} (H-E) {\cal A} | \Phi \chi \rangle =0.
\end{equation}
This results in a rather complicated equation for $|\chi\rangle$, which is possible to solve only by using 
strong approximations on 
$|\Phi\rangle$ and on the interaction of the particles. In a typical example $|\Phi\rangle$ describes fermions in harmonic oscillator 
potential wells located at some distance apart and $|\chi\rangle$ is the relative motion of the oscillator wells. It is easy to see that
if  $|\chi\rangle$ is expressed in terms of harmonic oscillator states, some of the lowest states in the relative motion space 
are not allowed due to the Pauli principle.

The fish-bone model is a model for the relative motion $|\chi\rangle$ \cite{schmid1}.
It is defined by an effective Hamiltonian 
\begin{equation}
h_{p}= h^{0}_{p}+v_{p}-\sum_{i,j}| u_{p,i}\rangle \langle u_{p,i}|(h^{0}_{p}+v_{p}-\epsilon_{p,i})|u_{p,j}\rangle \bar{M}_{p,ij}\langle u_{p,j} |,
\label{fishh}
\end{equation}
where $p$ refers to the angular momentum channel, $h^{0}_{p}$ is the kinetic energy operator and
$v_{p}$ is a local potential.  
Our knowledge about the internal structure $|\Phi\rangle$  and on the Pauli principle are incorporated in the last term.
The states $|u_{p,i}\rangle$ are eigenstates of the norm operator,
\begin{equation}
\langle \Phi \vec{r} | {\cal A} | \Phi u_{p,i} \rangle = (1-\eta_{p,i})\langle \vec{r} |u_{p,i}\rangle,
\end{equation}
where $\vec{r}$ is the center of mass distance of the two clusters. If the relative motion is forbidden by Pauli principle then $\langle \Phi \vec{r} | {\cal A} | \Phi u_{p,i} \rangle=0$,
and $\eta_{p,i}=1$. The $\eta_{p,i}$ eigenvalues are ordered such that $|\eta_{p,i}| \ge |\eta_{p,i+1}|$. 
The matrix $\bar{M}$ is then given by
\begin{equation}
\bar{M}_{p,ij}=
\begin{cases}
1-\frac{ \displaystyle 1-\eta_{p,i}}{ \displaystyle [(1-\bar{\eta}_{p,i}) (1-\bar{\eta}_{p,i})]^{1/2}}, & \text{if $i \le j$}, \\
 1-\frac{ \displaystyle  1-\eta_{p,j}}{ \displaystyle   [(1-\bar{\eta}_{p,j}) (1-\bar{\eta}_{p,i})]^{1/2}}, & 
 \text{if $i > j$},
 \end{cases}
\end{equation}
where $\bar{\eta}_{p,i}=0$ if $\eta_{p,i}=1$ and $\bar{\eta}_{p,i}=\eta_{p,i}$ otherwise.  
If the system  has only one Pauli-forbidden sate, the matrix $\bar{M}$, which exhibits a fish-bone-like structure, is given by
\begin{equation}
\bar{M}_{p}=\left(\begin{matrix}
1 & 1 & 1 & 1 & \ldots \\
1 & 0 & 1-\sqrt{\frac{1-\eta_{p,2}}{1-\eta_{p,3}}} & 1-\sqrt{\frac{1-\eta_{p,2}}{1-\eta_{p,4}}} &  \ldots \\
1 & 1-\sqrt{\frac{1-\eta_{p,2}}{1-\eta_{p,3}}} & 0 & 1-\sqrt{\frac{1-\eta_{p,2}}{1-\eta_{p,4}}} &   \ldots \\ 
1 & 1-\sqrt{\frac{1-\eta_{p,2}}{1-\eta_{p,3}}} & 1-\sqrt{\frac{1-\eta_{p,2}}{1-\eta_{p,4}}} & 0 &   \ldots \\ 
\vdots & \vdots & \vdots &\vdots & \ddots  \\
\end{matrix}\right).
\end{equation}

We can see that in the Hamiltonian (\ref{fishh}) the matrix elements of $h^{0}_{p}+v_{p}$ are eliminated  or partly suppressed 
by the fish-bone term. If a state $|u_{p,i}\rangle$ is fully Pauli forbidden, then the corresponding elements of $h^{0}_{p}+v_{p}$ are 
removed and from $h_{p}$. Consequently,  $|u_{p,i}\rangle$ becomes a solution
of the Schr\"odinger equation at zero energy. Or, if we take  $\epsilon_{p,i}$ nonzero in $h_{p}$, for Pauli forbidden states only, the 
corresponding $|u_{p,i}\rangle$ become solutions at $\epsilon$ energy.
In the fish-bone model   we take $\epsilon$ as a large positive number. Then the states at physically accessible energies 
become orthogonal to the Pauli forbidden states. If the Pauli-forbidden state is like a ground state harmonic oscillator
wave function, i.e.\ it is without any node, then the physical states has to be orthogonal to the Pauli forbidden state and 
must have a node.
So, the fish-bone model simulates the Pauli 
blocking by a node in the wave function at short distances.

\section{The fish-bone model for the $n-\alpha$ and $p-\alpha$ interactions}

We adopt a model that in the $\alpha$ particles the nucleons are in $0s$ states in a harmonic oscillator well of width parameter $a$. The norm kernel eigenvalues $|u_{p,i} \rangle$ are also harmonic oscillator states with the same width parameter \cite{zaikin}.
The eigenvalues $\eta_{p,i}$ in the $\mbox{}^{2}S_{1/2}$ channel are given by $1, 1/16, 1/16^{2}, \ldots$, and in the 
$\mbox{}^{2}P_{1/2}$
and $\mbox{}^{2}P_{3/2}$ channels are given by $-1/4, -1/(4 *16), -1/(4*16)^{2}, \ldots $. So, we have only one 
fully Pauli-forbidden sate in the $\mbox{}^{2}S_{1/2}$ channel. The other states are partly Pauli forbidden.
 We used the experimental phase shift compilation from Ref.\ \cite{satchler}.

The fish-bone model results in a Coulomb-like potential with non-local terms. We solved the equations by using the method 
of Ref.\ \cite{pzprc}. In this method the problem is written in a Lippmann-Schwinger integral equation form 
and the short-range terms are expanded in Coulomb-Sturmian basis. 
For the $\epsilon$ parameter of the fish-bone model, which aims  to remove the Pauli-forbidden states, we took 
$\epsilon=60000\:\mbox{MeV}$. In this range of $\epsilon$, the dependence of the results was beyond the fifth significant digit. 

The proton and neutron are spin $1/2$ particles and the proton has a charge. Therefore we seek the
 local part of the potential as a sum of a smeared Coulomb, a central  and a spin-orbit term
\begin{eqnarray}\label{locpot0}
v(r) &=& \frac{Z e^{2}}{r} \mbox{erf}\left( \sqrt{\frac{2a}{3}} r \right) + V(r)   \\ \nonumber
            &&+ [j(j+1)-s(s+1)-l(l+1)] \frac{v_{so}}{r} \frac{d}{dr} V(r),
\end{eqnarray}
where $s$ is the spin, $l$ is the orbital angular momentum, $j$ is the total angular momentum,
and $Z=0$ for the $n-\alpha$ system and $Z=2$ for the $p-\alpha$ system.

\begin{figure}[!ht]
\centering
\includegraphics[width=7.5cm]{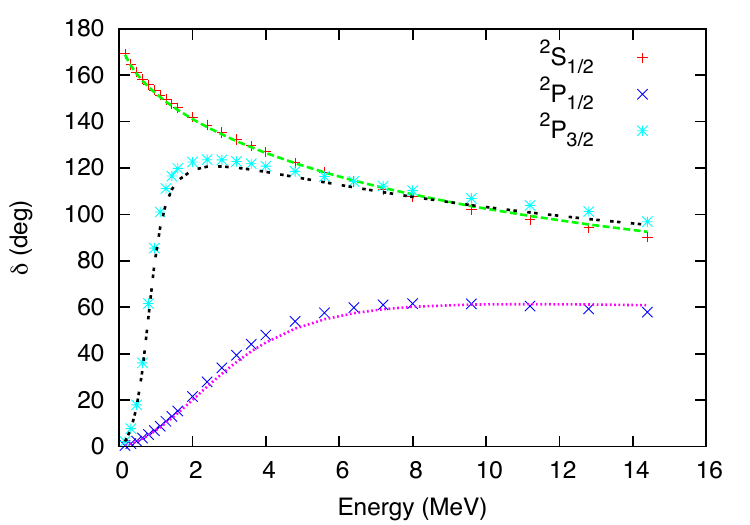}
\caption{Fit to the experimental  $\mbox{}^{2}S_{1/2}$, $\mbox{}^{2}P_{1/2}$, and  $\mbox{}^{2}P_{3/2}$ $n-\alpha$ phase shifts.}
\label{fig1}
\end{figure}

\begin{figure}[!ht]
\centering
\includegraphics[width=7.5cm]{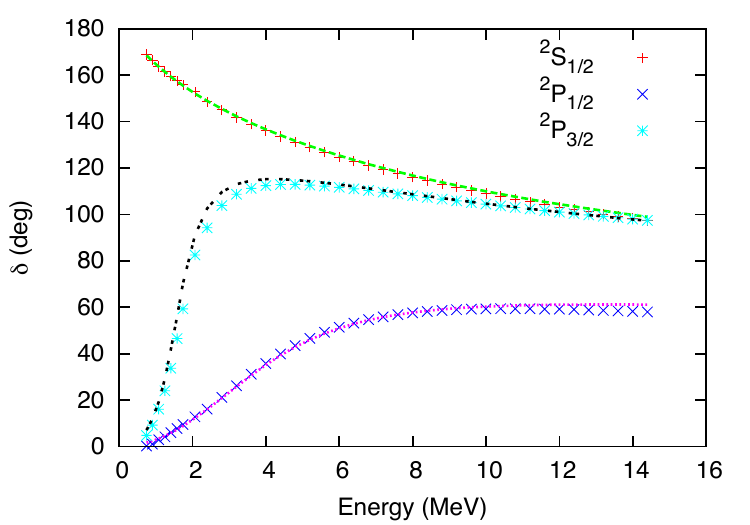}
\caption{Fit to the experimental  $\mbox{}^{2}S_{1/2}$, $\mbox{}^{2}P_{1/2}$, and  $\mbox{}^{2}P_{3/2}$ $p-\alpha$ phase shifts. }
\label{fig2}
\end{figure}

\begin{figure}[!ht]
\centering
\includegraphics[width=7.5cm]{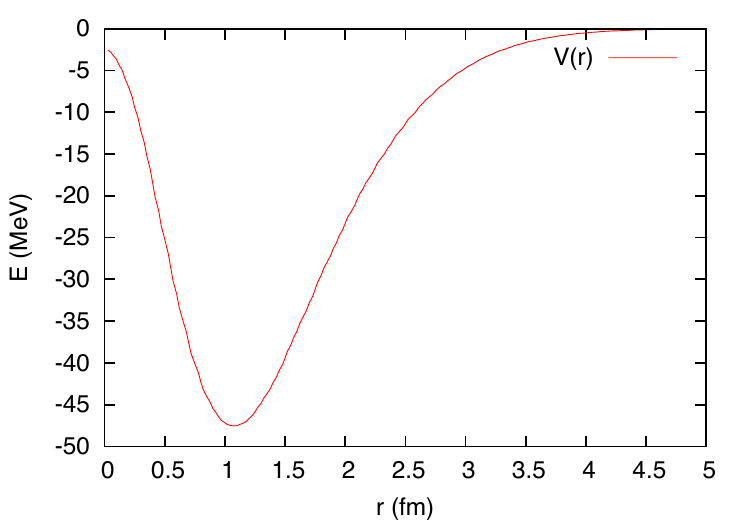}
\caption{The local potential $V(r)$. }
\label{fig3}
\end{figure}

We took the harmonic oscillator width parameter $a$ and the spin-orbit coupling term $v_{so}$ as fitting parameters and tried out 
several forms for $V(r)$. We achieved an excellent fit to the phase shift values with a double Gaussian potential 
\begin{equation}
V(r)= v_{1}\exp(-\alpha_{1} r^{2}) + v_{2}\exp(-\alpha_{2} r^{2}).
\end{equation}
The ''best fit'' parameters are $a= 0.214\:\mbox{fm}^{-2}$, $v_{1}= -84.068\:\mbox{MeV}$, 
$\alpha_{1}= 0.3199\:\mbox{fm}^{-2}$,  
$v_{2}= 81.66\:\mbox{MeV}$, $\alpha_{2}= 1.7719\:\mbox{fm}^{-2}$, and $v_{so}=  -0.28497 $.
Our results for the $n-\alpha$ and $p-\alpha$ scattering are given in Figures \ref{fig1} and \ref{fig2}, respectively, and $V(r)$ is shown in Fig.\ \ref{fig3}.

We can see that the $\mbox{}^{2}S_{1/2}$ phase shifts start at $180^{\circ}$, although neither the
$n-\alpha$ nor the $p-\alpha$ system has any bound state. This seems to be in
contradiction to the Levinson theorem, which connects the zero-energy phase shift to the number of bound states. 
The bound state, which is required by the low energy behavior of the
$\mbox{}^{2}S_{1/2}$ phase shift, is forbidden by the Pauli principle. Although the Pauli-forbidden states are 
absent from the spectrum, their effects are clearly visible in the phase shift.

\section{Summary and conclusion}

In this work we propose a new parametrization for the $n-\alpha$ and $p-\alpha$ fish-bone potentials.  
We determined the parameters of the potential by fitting to the experimental phase shifts. We found that if we incorporate 
our knowledge on the charge, the spin and the composite structure into the form of the potential, 
then the potential to be fitted is really very simple.
With the same set of parameters, six parameters altogether, and without any angular momentum dependence in $V$, 
we achieved a very good description for
the $n-\alpha$ and $p-\alpha$ low energy scattering data for all the relevant partial waves. 
In a forthcoming work we
will study this potential in the $\alpha-\alpha-n$ and $\alpha-\alpha-p$ systems.

We believe that the fish-bone model deserves further attention.
We can see that the fish-bone model really 
provides a good account for the composite structure of the constituents and of the Pauli principle. Then the local potential, 
which is to be determined by a fitting procedure, becomes very simple. We can also conclude that the strong angular dependence 
of the potentials may be mainly due to inadequate treatment of the internal structure of the composite particles.
The fish-bone model uses the concept of partly Pauli forbidden states as well. So, it can model Pauli effect even if there is no
complete Pauli blocking, like in the nucleon-nucleon potential.


\begin{thebibliography}{00}

\bibitem{schmid1} E.~W.~Schmid,  Z.\ Phys.\ A.\  {\bf 297}, 105 (1980).

\bibitem{schmid2}   E.~W.~Schmid,  Z.\ Phys.\ A.\   {\bf  302}, 311 (1981). 

\bibitem{kircher} R.~Kircher and E.~W.~Schmid,  Z.\ Phys.\ A.\  {\bf 299} 241 (1981).

\bibitem{oryu} S.~Oryu and H.~Kamada, Nucl.\ Phys.\ A.\ {\bf 493} 91 (1989).

\bibitem{our3a}  J.~P.~Day, J.~E.~McEwen, M.~Elhanafy, E.~Smith, R.~Woodhouse, and Z.~Papp,
Phys.\ Rev.\ C {\bf 84}, 034001 (2011). 

\bibitem{hahn} K.~Hahn, E.~W.~Schmid, and P.~Doleschall, Phys.\ Rev.\ C {\bf 31}, 325 (1985).

\bibitem{zaikin} D.~A.~Zaikin, Nucl.\ Phys.\ {\bf A357}, 584 (1971).

\bibitem{satchler} G.\ R.\ Satchler, L.\ W.\ Owen, A.\ J.\ Elwyn, G.\ L.\ Morgan, and R.\ L.\ Walter,
Nucl.\ Phys.\ {\bf A112}, 1 (1968).

\bibitem{pzprc} Z.\ Papp, Phys.\ Rev.\ C {\bf 38}, 2457 (1988).

\end{thebibliography}
\end{document}